\input{aipcheck}

\documentclass[
    ,final            
  ]
  {aipproc}

\usepackage[]{graphicx}

\layoutstyle{8x11single}

\newcommand{\be}{\begin{equation}}
\newcommand{\ee}{\end{equation}}
\newcommand{\bea}{\begin{eqnarray}}
\newcommand{\eea}{\end{eqnarray}}

\begin{document}

\title{Exotic-singularity-driven dark energy}


\author{Mariusz P. D\c{a}browski and Tomasz Denkiewicz}{
  address={Institute of Physics, University of Szczecin, Wielkopolska 15, 70-451 Szczecin, Poland}
}



\begin{abstract}
We discuss various types of exotic (non-standard) singularities in the Universe:
a Big-Rip (BR or type I), a Sudden Future Singularity (SFS or type II), a Generalized Sudden Future Singularity,
a Finite Scale Factor singularity (FSF or type III), a Big-Separation (BS or type IV) and a $w$-singularity. They are characterized by violation of all or some of the energy conditions which results in a blow-up of all or some of the physical quantities: the scale factor, the energy density, the pressure, and the barotropic index. We relate the emergence of these singularities with physical theories (superstring, brane, higher-order gravity, loop quantum cosmology). We show how the models involving exotic singularities may serve as dark energy by applying the observational data. In particular, we show that some of these exotic singularities (though being of a weak type
according to relativistic definitions) may occur in the near future of the universe.
\end{abstract}

\maketitle


\section{Introduction.}

The paradigm of the standard evolution of the universe from a Big-Bang to a Big-Crunch or to an asymptotic emptiness has been broken up once the possibility to admit supernegative pressure matter (phantom) in the universe appeared. Since then cosmologists started studying more exotic behavior of the cosmological models with respect to their asymptotic properties both in the past and in the future. In this contribution we will discuss such options and try to relate them to the observational data.

\section{Standard Big-Bang/Big-Crunch versus exotic singularities.}

It is well-known that the standard Einstein-Friedmann equations are the two equations for the three unknown functions of time: the scale factor $a(t)$, the pressure $p(t)$, and the energy density $\varrho(t)$ ($K=0, \pm 1$)\\
\bea
\label{rho}
\varrho &=& 3 \left(\frac{\dot{a}^2}{a^2} + \frac{K}{a^2}
\right)~,\\
\label{p}
p &=& - \left(2 \frac{\ddot{a}}{a} + \frac{\dot{a}^2}{a^2} + \frac{K}{a^2}
\right)~.
\eea
In order to make the system closed, one usually assumes additionally {\it an equation of state}, e.g., of a barotropic type:
\be
\label{eos}
p(t) = w\varrho(t)~,
\ee
which allows to solve for $a(t)$. Until late nineties of the twentieth century most of the cosmologists studied only the three simplest, say ``standard'', solutions with $w=$ const. $\geq -1$. Each of them starts with a Big-Bang singularity in which $a \to 0$, $\varrho \to \infty$, $p \to \infty$. One of these solutions (of $K=+1$) terminates at a second singularity (a Big-Crunch) where $a \to 0$, $\varrho \to \infty$, $p \to \infty$. The other two solutions ($K=0, -1$) continue to an asymptotic emptiness $\varrho \to 0$, $p \to 0$ for $a \to \infty$. The Big-Bang and the Big-Crunch singularities are the strong singularities according to general relativity since they exhibit {\it a geodesic incompletness} and {\it a curvature blow-up}. One of the relativistic paradigms then was the obedience of the strong energy condition (SEC)
\bea
R_{\mu\nu} V^{\mu} V^{\nu} &\geq& 0, \hspace{0.5cm} V^{\mu} - {\rm a}\hspace{5pt}
{\rm timelike}\hspace{5pt} {\rm  vector}~,
\eea
($R_{\mu\nu}$ - Ricci tensor) which in terms of the energy density
and pressure is equivalent to
\bea
\label{strong}
\varrho + 3p \geq 0, \hspace{0.5cm} \varrho + p \geq 0~.
\eea
From (\ref{rho}) and (\ref{p}) one has
\bea
\label{accel}
- \frac{4\pi G}{3} (\varrho + 3p) &=& \frac{\ddot{a}}{a} = - q
H^2~,\hspace{0.2cm} (q = -\frac{\ddot{a}a}{\dot{a}^2}; \hspace{0.1cm} H =
\frac{\dot{a}}{a})~,
\eea
which together with (\ref{strong}) means that
\bea
\ddot{a} &\leq& 0, \hspace{0.2cm} {\rm or} \hspace{0.2cm} q \geq 0~,
\eea
and so the universe was supposed to decelerate its expansion in this
``standard'', or better, ``pre-supernovae'' case.

The seminal works on supernovae \cite{supernovaeold} showed that the best-fit model was for
$q_0 = \Omega_{M}/2 - \Omega_{\Lambda} < 0$, so that $\ddot{a}>0$ and this favored matter
with negative pressure ($-1 \leq w \leq 0$) as dominating the universe - later this type of
matter was dubbed the dark energy or quintessence. The discovery gave evidence for the strong energy condition (\ref{strong}) violation, but the paradigm of the appearance of the ``standard'' Big-Bang/Big-Crunch singularities remained untouched.

\begin{figure}[h]
\caption{Cosmic ``no-hair'' theorem.}
\includegraphics[width=10cm]{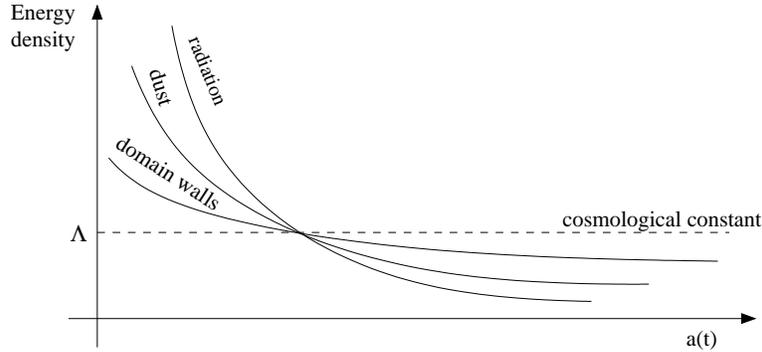}
\label{nohair}
\end{figure}

This is no wonder in view of the cosmic ``no-hair'' theorem which says that if only $w =$ const. $\geq -1$ matter appears in the universe (so merely the strong energy condition is
violated), the cosmological constant ($w=-1$) of any small fraction will always dominate
the future evolution of the universe (Fig. \ref{nohair}). Then, any combination of the dark
energy with $w=$ const. $\geq -1$ leads to ``standard'' Big-Bang/Big-Crunch cosmological
singularities or, eventually,  to the emptiness described asymptotically by the de Sitter space.

However, most recent data \cite{supernovaenew} from WMAP, SDSS and supernovae gave a combined bound on the value of the barotropic index $w$ and showed that there was no sharp cut-off of the data at $p = -\varrho$! This meant that the dark energy with $p < -\varrho$ could also be admitted. This type of the dark energy was named phantom. Now, we even have more accurate data. Knop et al. \cite{knop} gives a combined bound on the $w$-index which comes from supernovae, CMB and 2dFGRS which is  $w = -1.05^{+0.15}_{- 0.20}$ (statistical) $\pm 0.09$ (systematic) and even more recently Kowalski et al. \cite{kowalski} analyzed 307 supernovae plus baryon acoustic ascillations (BAO) and CMB to get $w = -1.001^{+ 0.059}_{-0.063}$ (statistical) $^{+0.063}_{-0.066}$ (systematic).

This can be considered as the evidence for the cosmic ``no-hair'' theorem violation in the
sense that even a small and still admissible fraction of phantom dark energy will dominate the future evolution
of the universe and it does not finish its evolution in either a Big-Crunch or the emptiness.

\section{Phantom dark energy. Big-Rip (type I) as an exotic singularity.}

Phantom is dark energy of {\it a very large negative pressure} \cite{phantom}
\bea
p < - \varrho, \hspace{0.5cm} {\rm or} \hspace{5pt} w<-1~,
\eea
which besides SEC, also violates the remained energy conditions, i.e., the null (NEC)
\bea
\label{null}
T_{\mu\nu} k^{\mu} k^{\nu} &\geq& 0, \hspace{0.5cm} k^{\mu} - {\rm
a}\hspace{5pt}{\rm null}\hspace{5pt} {\rm vector}~, \hspace{5pt}
{\rm i.e.,} \hspace{0.5cm} \varrho + p \geq 0~,
\eea
the weak (WEC)
\bea
\label{weak}
T_{\mu\nu} V^{\mu} V^{\nu} &\geq& 0, \hspace{0.2cm} V^{\mu} - {\rm
a}\hspace{5pt}{\rm timelike}\hspace{5pt} {\rm  vector}~,\hspace{5pt}
{\rm i.e.,} \hspace{0.1cm} \varrho + p \geq 0, \hspace{0.1cm} \rho \geq 0~,
\eea
and the dominant energy (DEC)
\bea
\label{dominant}
T_{\mu\nu} V^{\mu} V^{\nu} &\geq& 0, \hspace{0.2cm} T_{\mu\nu} V^{\mu} -
{\rm not}\hspace{5pt} {\rm  spacelike}~,\hspace{5pt}
{\rm i.e.,} \hspace{0.1cm} \mid p \mid \leq \varrho, \hspace{0.1cm}
\varrho \geq 0~.
\eea
The most interesting point about phantom can be made clear, if we take
\be
\mid w + 1 \mid = - (w+1) > 0~,
\ee
so that the conservation law for phantom gives
\be
\varrho \propto a^{3 \mid w+1 \mid}~.
\ee
It then appears that once the universe grows bigger, it also becomes denser, and it
finally becomes dominated by phantom which overcomes the $\Lambda$-term. What is even more
surprising, an exotic future singularity appears which is now called a Big-Rip in which
$\varrho \to \infty$, $p \to \infty$ for $a \to \infty$. It appears that curvature invariants $R^2$, $R_{\mu\nu}R^{\mu\nu}$, $R_{\mu\nu\rho\sigma}R^{\mu\nu\rho\sigma}$ diverge at a Big-Rip.
Deeper studies show that only for $-5/3 < w < -1$ the null geodesics are geodesically {\it complete}; for other values of $w$, including all timelike geodesics, there is a geodesic {\it incompleteness} \cite{lazkoz06} - the singularity is reached in a finite proper time.
This means that a Big-Rip is a serious singularity and cannot be passed thorough.

\begin{figure}[h]
\label{phduality}
\caption{Phantom duality.}
\scalebox{0.5}{\includegraphics[width=14cm]{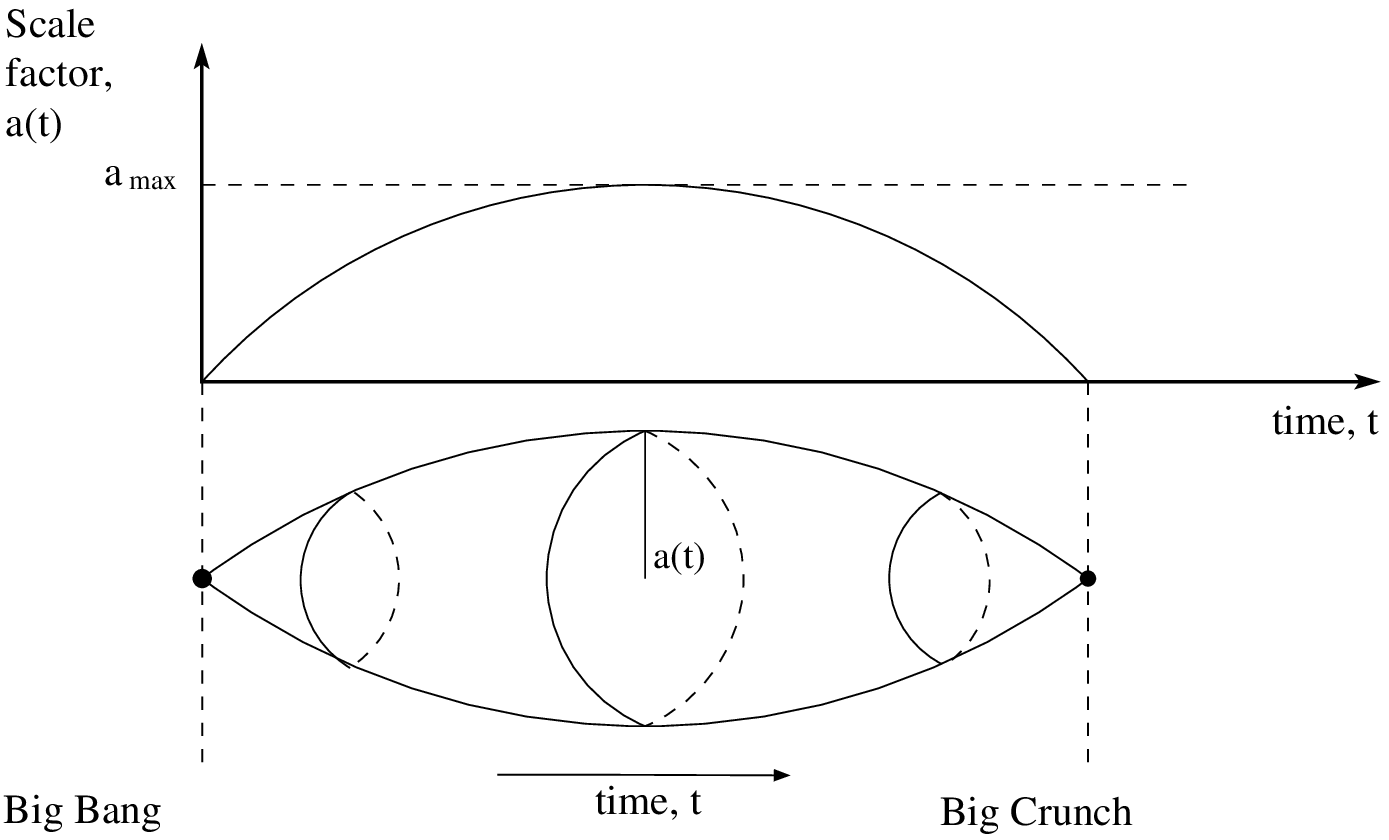}\includegraphics[width=14cm]{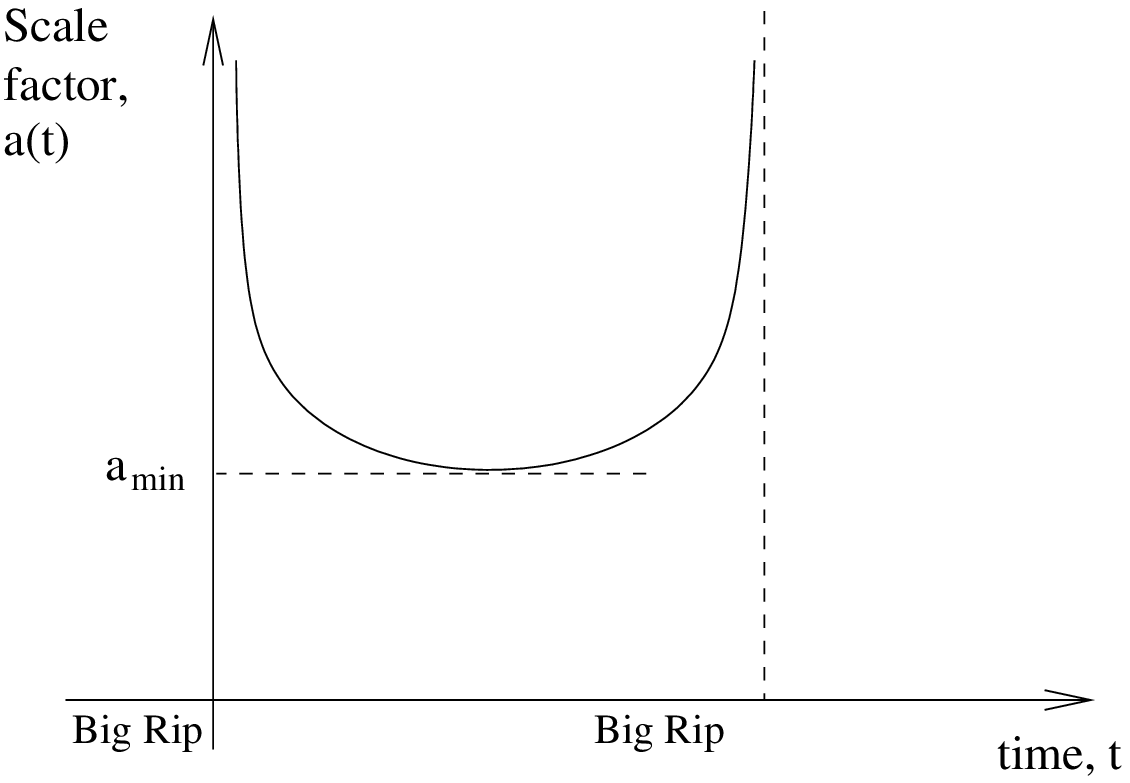}}
\end{figure}

Besides, there is an interesting duality between the standard matter ($p>-\varrho$)
and phantom ($p<-\varrho$) which is like the scale factor duality in superstring cosmology
\cite{meissner} (see Fig. \ref{phduality}) \\
\be
w \leftrightarrow - (w+2) \hspace{1.cm} {\rm or} \hspace{0.2cm} {\rm
  better} \hspace{0.2cm} \gamma \leftrightarrow - \gamma~, \hspace{0.2cm} \gamma = w+1
\ee
i.e.
\be
a(t) \leftrightarrow \frac{1}{a(t)}
\ee

According to what has been said at the beginning of the section, the cosmological bounds
allow phantom as a source of dark energy, so that we can talk either about phantom or
a Big-Rip-singularity-driven dark energy. In such a scenario, the universe begins with
a Big-Bang and terminates at a Big-Rip so that standard Hot-Big-Bang cosmology is preserved.
Besides, there is a turning point at $z = 0.46$ (since the jerk parameter $j_0 > 0$)
\cite{supernovaenew}. A comparison with the data also shows that a Big-Rip is
possible in the far future from now, i.e., in about $t=t_{BR} \approx$ 20 Gyr.

\section{Sudden Future Singularity (type II) as an exotic singularity.}

Strange properties of a Big-Rip gave a push to studies of some other exotic
(or just ``non-standard'') types of singularities and the sources of dark energy which
can be related to them. One of such singularities is a Sudden Future Singularity (SFS)
(or type II \cite{nojiri}) which manifests as a singularity of pressure (or $\ddot{a}$)
only and which leads to the dominant energy condition violation only. The hint how to get
it was given by Barrow \cite{barrow04}. Namely, one has to release the assumption about the
imposition of an equation of state
\be
p \neq p(\varrho),
\ee
onto the set of equations (\ref{rho})-(\ref{p}) which allows the evolution of the energy
density and pressure being unconstrained. Then, one may choose a special form of the scale
factor (which, in fact, is motivated by fundamental theories such as superstring theory) as:
\be
\label{sf2} a(t) = a_s \left[\delta + \left(1 - \delta \right) y^m -
\delta \left( 1 - y \right)^n \right]~, \hspace{0.5cm} y \equiv
\frac{t}{t_s}
\ee
where $a_s \equiv a(t_s)=$ const. and $\delta, m, n =$ const. Obviously, for $t=0$, one has a Big-Bang singularity. The derivatives of the scale factor are
\bea
\label{dota}
\dot{a} &=& a_s \left[ \frac{m}{t_s} \left(1 - \delta \right)y^{m-1} + \delta\frac{n}{t_s}\left( 1 - y \right)^{n-1} \right]~,\\
\label{ddota}
\ddot{a} &=& \frac{a_s}{t_s^2} \left[ m\left(m-1\right) \left(1 -
    \delta \right) y^{m-2} - \delta n \left(n-1 \right) \left(1-y \right)^{n-2} \right]~.
\eea
If we assume that
\be
\label{nq}
1 < n < 2,
\ee
then, using Einstein equations (\ref{rho})-(\ref{p}), we get the following conditions
\be
a = {\rm const.}, \hspace{0.5cm} \dot{a} = {\rm const.}, \hspace{0.5cm} \varrho = {\rm const.}, \hspace{0.5cm} \ddot{a} \to \mp \infty \hspace{0.5cm} p \to \pm \infty \hspace{0.5cm} {\rm
   for } \hspace{0.5cm} t \to t_s~~.
\ee
One can show that the standard Friedmann cosmology limit (which has an imposed equation of state) is easily obtained in the limit of ``nonstandardicity'' parameter $\delta \to 0$.
The parameter $m$ can be taken to be just a form of the $w$ parameter present in the barotropic equation of state:
\begin{center}
-- $0 < m \leq 1$ when $w \geq -1/3$ (standard matter);\\
-- $m>1$ when $-1 < w < -1/3$ (quintessence);\\
-- $m<0$ when $w < -1$ (phantom).
\end{center}
Near to an SFS one has
\be
a_{SFS} = a_s \left[1 - m(1 - \delta) (1 - y) \right]~ .
\ee
An important point of our considerations is that unless we take $m>1$ or $m<0$ as an independent source of the dark energy, then the acceleration is possible only for $\delta < 0$, so that we will call it pressure-driven dark energy (whatever it is!). Now, it emerges that unlike the Big-Bang or the Big-Rip, an SFS is a ``weak'' singularity, since it is determined by a blow-up of the Riemann tensor and its derivatives but not the scale factor (or its inverse) itself.

In fact, individual geodesics do not feel an SFS at all, since geodesic
equations are not singular for $a_s = a(t_s)=$ const. \cite{lazkoz06}
\bea
\left(\frac{dt}{d\tau}\right)^2 &=& A + \frac{P^2 +
KL^2}{a^2(t)}~,\\
\frac{dr}{d\tau} &=& \frac{P_1 cos{\phi} + P_2
\sin{\phi}}{a^2(t)} \sqrt{1-Kr^2}~,\\
\frac{d\phi}{d\tau} &=& \frac{L}{a^2(t)r^2}~,
\eea
with $\tau-$ a proper time, $A, P, K, L =$ const. However, the geodesic deviation equation
\bea
\frac{D^2n^{\alpha}}{d\lambda^2} + R^{\alpha}_{~\beta \gamma
\delta} u^{\beta} n^{\gamma} u^{\delta} = 0~,
\eea
feels an SFS since at $t=t_s$ we have a blow-up of the Riemann tensor $R^{\alpha}_{~\beta \gamma \delta} \to \infty$. We can summarize the features of sudden singularities as follows:
\begin{itemize}

\item There is no geodesic incompletness in the sense that one can extend the evolution behind the singularity and so SFSs are not the final state of the universe;

\item Point particles do not even see SFSs, while extended objects may suffer instantaneous infinite tidal forces, though still may not be crushed;

\item They are {\it weak} curvature singularities i.e. in-falling observers or detectors are not destroyed by tidal forces. In fact, cosmic strings may pass through an SFS and are not destroyed at all \cite{strings};

\item According to Tipler \cite{tipler} a singularity is weak if the double integral\\
   $\int_0^{\tau} d\tau' \int_0^{\tau'} d\tau'' R_{ab}u^a u^b$ \\
   does not diverge on the approach to it at $\tau = \tau_s$ and this happens for an SFS;

\item According to Kr\'olak \cite{krolak} a singularity is weak if the single integral \\
   $\int_0^{\tau} d\tau' R_{ab}u^a u^b$ \\
   does not diverge on the approach to a singularity at $\tau = \tau_s$, and this also happens for an SFS;

\item A singularity may be weak according to Tipler, but strong according to Kr\'olak;

\item There is a conclusion that a Big-Rip and an SFS are of different nature;

\item A Big-Rip (similar as a Big-Crunch) is really {\it the final state} of
  the evolution of the universe, though for $a \to \infty$ the 4-velocity $u^{\mu} \to
  0$ (particles gradually slow their movement on a coordinate chart).
\end{itemize}

\section{Type III (Finite Scale Factor) exotic singularity.}

Type III singularities \cite{nojiri}, which we will call Finite Scale Factor (FSF) singularities, are characterized by the following conditions:\\
\be
a=a_s = const., \hspace{0.5cm} \varrho \to \infty, \hspace{0.5cm} \dot{a}_s \to
  \infty, \hspace{0.5cm} \vert p \vert \to \infty, \hspace{0.5cm} \ddot{a}_s \to \infty~~.
\ee
The simplest way to obtain them is to apply the scale factor as given previously for an SFS by the formula (\ref{sf2}), i.e.,
\be
\label{sf22} a(t) = a_s \left[\delta + \left(1 - \delta \right) y^m -
\delta \left( 1 - y \right)^n \right]~, \hspace{0.5cm} y \equiv
\frac{t}{t_s}~~,
\ee
where $a_s \equiv a(t_s)=$ const. and $\delta,A, m, n =$ const.,
but with the range of the parameter $n$ changed from $1<n<2$ into
\be
0< n < 1 ~~.
\ee
FSF singularities appear to be weak according to the Tipler's definition, but strong
according to the Kr\'olak's \cite{lazkoz06}.

\section{Generalized Sudden Future, type IV (Big Separation) and $w-$singularities.}

Sudden future singularities may be generalized to generalized sudden future singularities (GSFS). In order to discuss them, one should consider a general scale factor time derivative of an order $r$ as follows:
\bea
\label{dotageneral} a^{(r)} &=&
a_s \left[ \frac{m(m-1)...(m-r+1)}{t_s^r} \left(1 - \delta \right)
y^{m-r} \right. \nonumber \\
&+& \left. (-1)^{r-1} \delta \frac{n(n-1)...(n-r+1)}{t_s^r} \left( 1
- y \right)^{n-r}\right]~.
\eea
By choosing $r-1 < n < r$ \cite{barrow04,lake}, for any integer $r$, we have a
singularity in the scale factor derivative $a^{(r)}$, and consequently in the appropriate pressure derivative $p^{(r-2)}$ as well as in the Hubble parameter time derivative $H^{(r-1)}$. Also, in such a case, the energy density $\varrho$ and the pressure $p$ remain finite. An amazing feature of this model is that it fulfills all the energy conditions for any $r \geq 3$!

Type IV singularity which we will call a Big-Separation is when:
\be
a=a_s = {\rm const.}, \hspace{0.5cm}\varrho \to 0, \hspace{0.5cm} p  \to 0, \hspace{0.5cm} a^{(r)} \to \infty, \hspace{0.5cm}  H^{(r-1)} \to \infty \hspace{0.5cm} (r \geq 3)
\ee
and so it is similar to a Generalized Sudden Future singularity with only one exception:
it also gives the divergence of the barotropic index in the barotropic equation of state
\be
p(t) = w(t) \varrho(t),
\ee
i.e.,
\be
w(t) \to \infty.
\ee

Another exotic is a $w-$singularity, or better a $w-$index singularity, which appears to be a singularity of a $w-$index only without the divergence of the higher-derivatives of the scale factor $a^{(r)}$ or any other physical quantity \cite{wsing}. In order to get it, we choose
\be
\label{SF1}
a(t)=A+B\left(\frac{t}{t_s}\right)^{\frac{2}{3\gamma}}+C\left(D-\frac{t}{t_s}\right)^n~,
\ee
where $A$, $B$, $C$, $D$, $\gamma$, $n$, and $t_s$ are constants and we impose the conditions:
\be
\label{cond1}
a(0)=0, \hspace{0.5cm} a(t_s)=const.\equiv a_s, \hspace{0.5cm} \dot{a}(t_s)=0, \hspace{0.5cm} \ddot{a}(t_s)=0~,
\ee
which leads to the following form of the scale factor:
\bea
\label{SFF}
a(t)&=&\frac{a_s}{1-\frac{3\gamma}{2}\left(\frac{n-1}{n-\frac{2}{3\gamma}}\right)^{n-1}}\nonumber
\\&+&\frac{1-\frac{2}{3\gamma}}{n-\frac{2}{3\gamma}}\frac{na_s}{1-\frac{2}{3\gamma}\left(\frac{n-\frac{2}{3\gamma}}{n-1}\right)^{n-1}}\left(\frac{t}{t_s}\right)^{\frac{2}{3\gamma}} \nonumber
\\&+& \frac{a_s}{\frac{3\gamma}{2}\left(\frac{n-1}{n-\frac{2}{3\gamma}}\right)^{n-1}-1}\left(1-\frac{1-\frac{2}{3\gamma}}{n-\frac{2}{3\gamma}}\frac{t}{t_s}\right)^n~,
\eea
with the admissible values of the parameters: $\gamma >0$ and $n \neq 1$.
By using the field equations (\ref{rho})-(\ref{p}), and writing an explicit equation of state we conclude that we have a blow-up of the effective barotropic index, i.e.,
\be
w(t_s) = \frac{1}{3}\left[2 q(t_s) - 1 \right] \to \infty~,
\ee
where $q(t_s)$ is a deceleration parameter at $t=t_s$, and
\be
p(t_s) \to 0; \hspace{0.3cm} \varrho(t_s) \to 0~,
\ee
which is equivalent to the condition (\ref{cond1}). Besides, there is an amazing duality between the Big-Bang and the $w$-singularity in the form
\be
\label{duality}
p_{BB} \leftrightarrow \frac{1}{p_w}, \hspace{0.3cm} \varrho_{BB} \leftrightarrow \frac{1}{\varrho_w}, \hspace{0.3cm} w_{BB} \leftrightarrow \frac{1}{w_w}~~.
\ee
In other words, one has the maps
\be
p_{BB} \to \infty; \hspace{0.3cm}\varrho_{BB} \to \infty; \hspace{0.3cm} w_{BB} \to 0; a_{BB} \to 0; \hspace{0.3cm} p_w \to 0; \hspace{0.3cm} \varrho_w \to 0; \hspace{0.3cm} w_w \to \infty; \hspace{0.3cm} a_w \to a_s= const.,
\ee
which are shown explicitly in the figures \ref{w1} and \ref{w2} below.

\begin{figure}[h]
\unitlength1cm
\caption{w-duality for the pressure and the energy density}
\scalebox{0.45}{\includegraphics[angle=0]{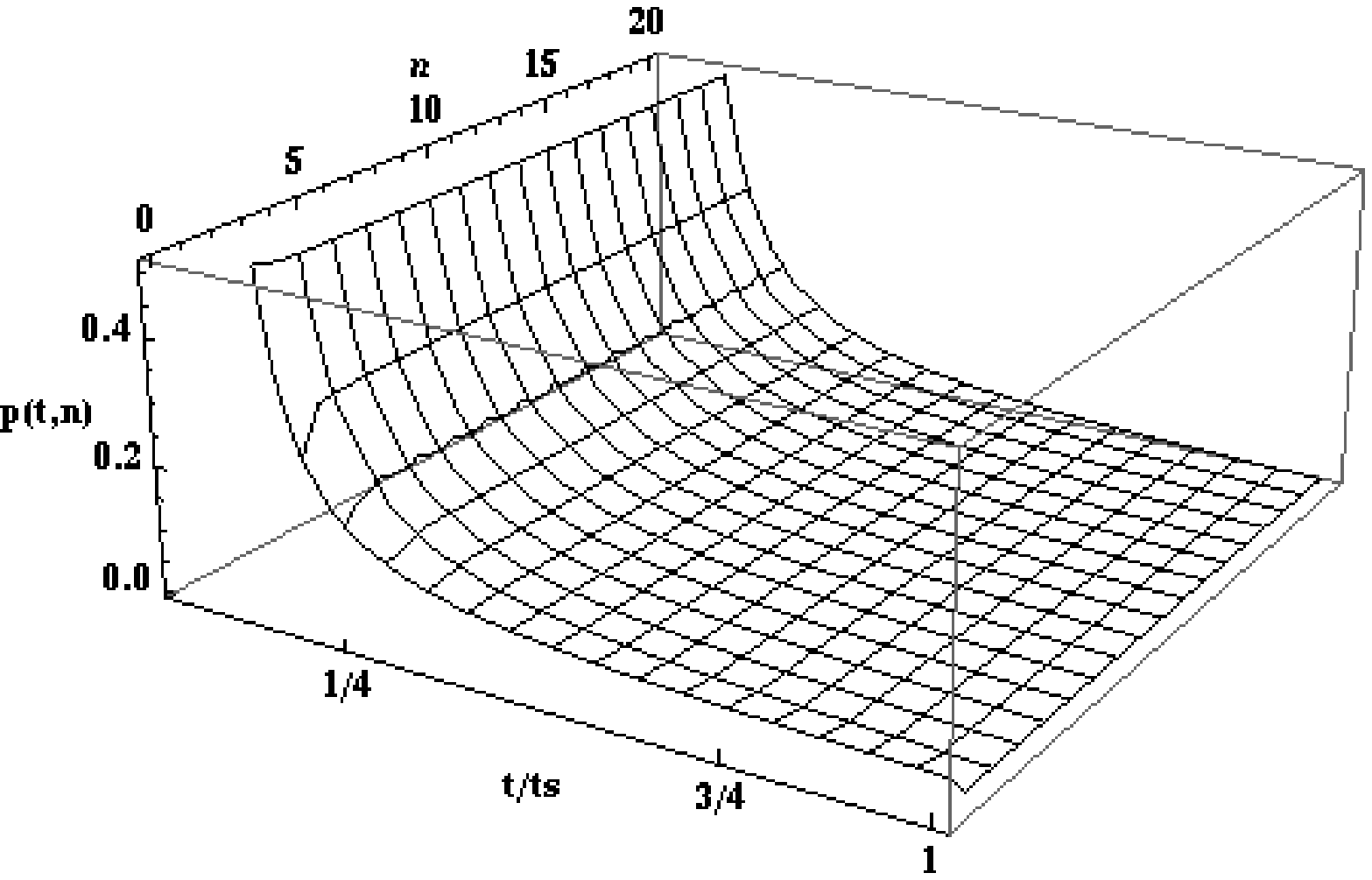}\includegraphics[angle=0]{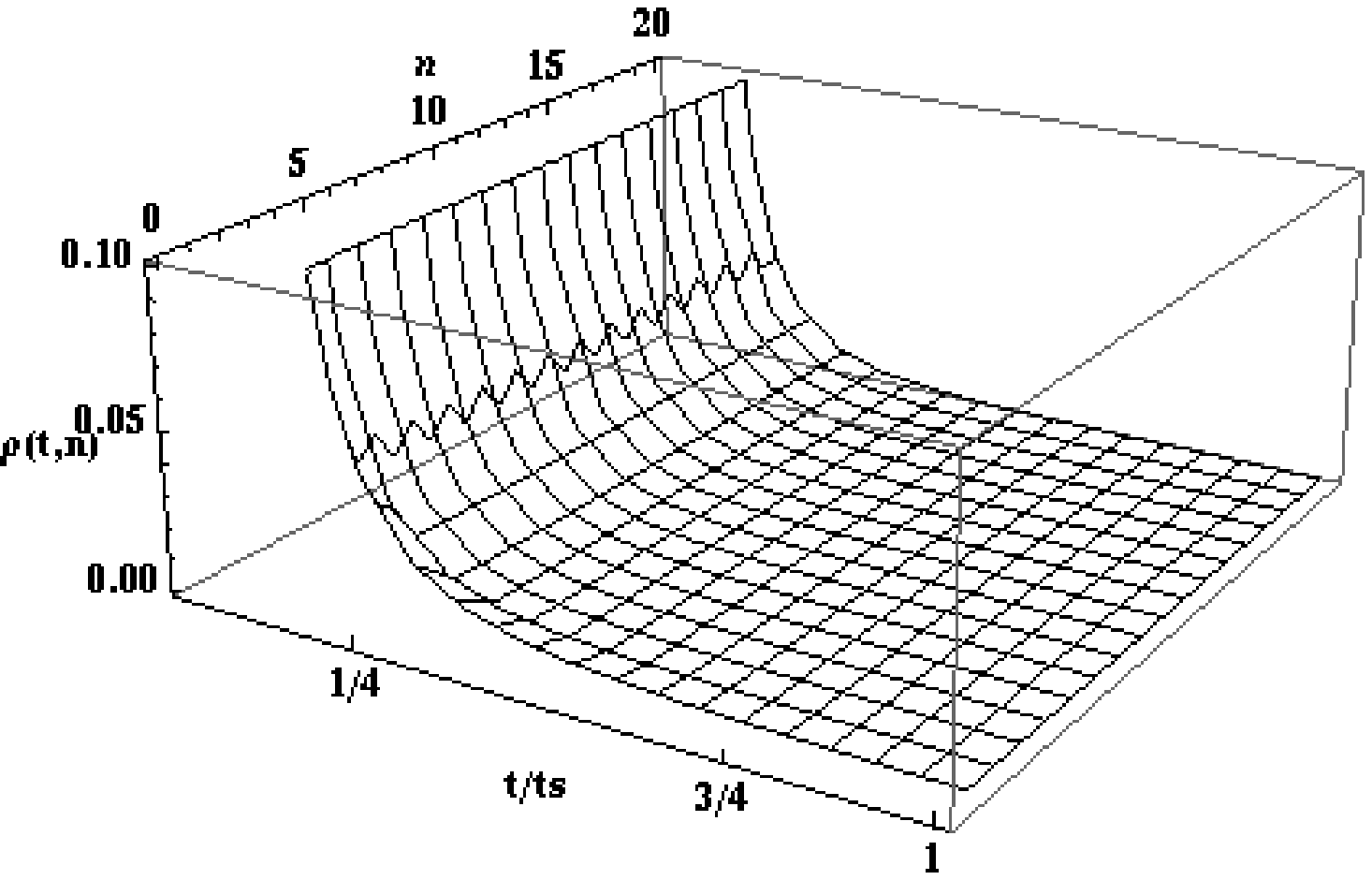}}
\label{w1}
\end{figure}

\begin{figure}[h]
\unitlength1cm
\caption{w-duality for the scale factor and the w-index}
\scalebox{0.45}{\includegraphics[angle=0]{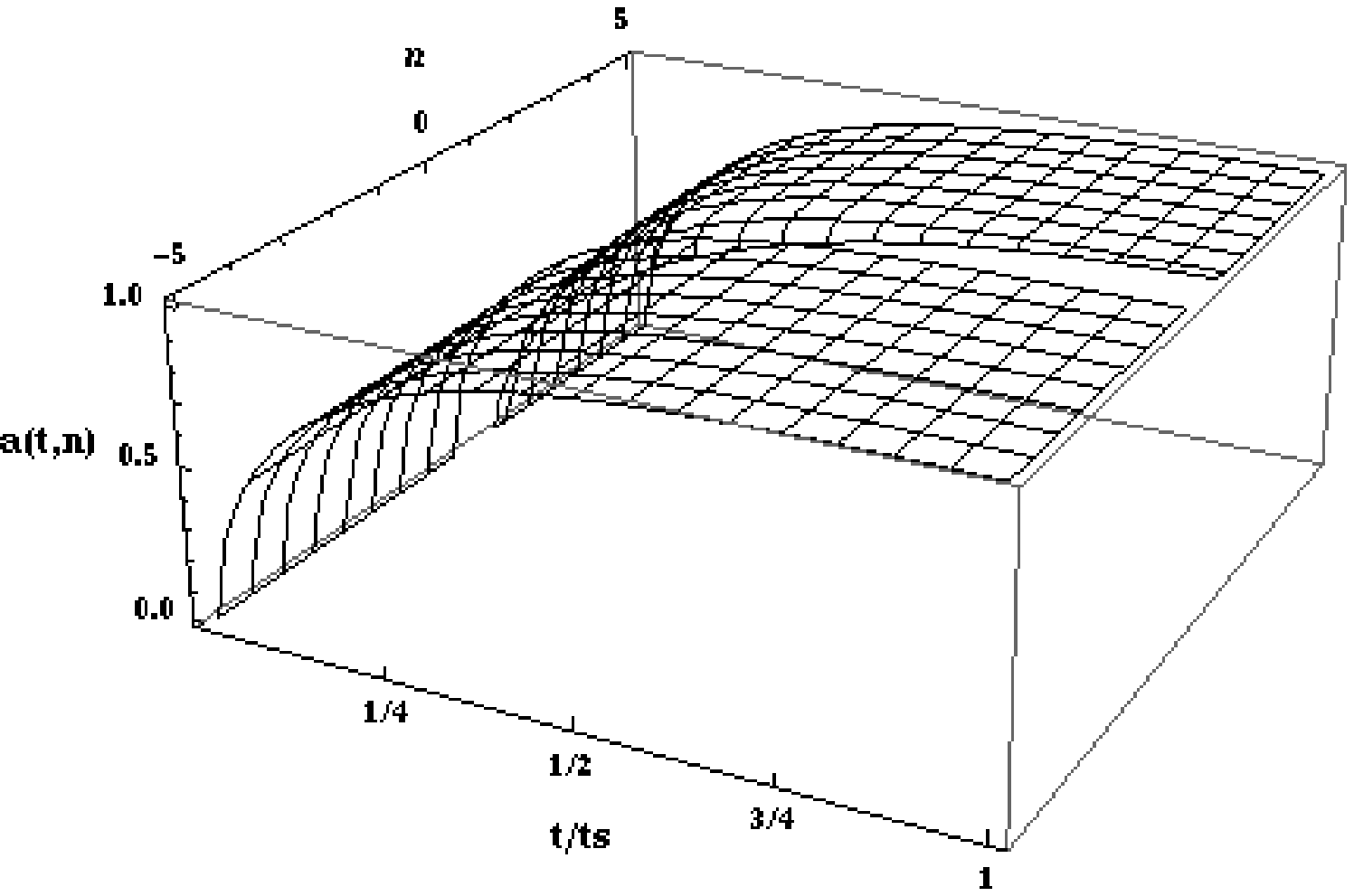}\includegraphics[angle=0]{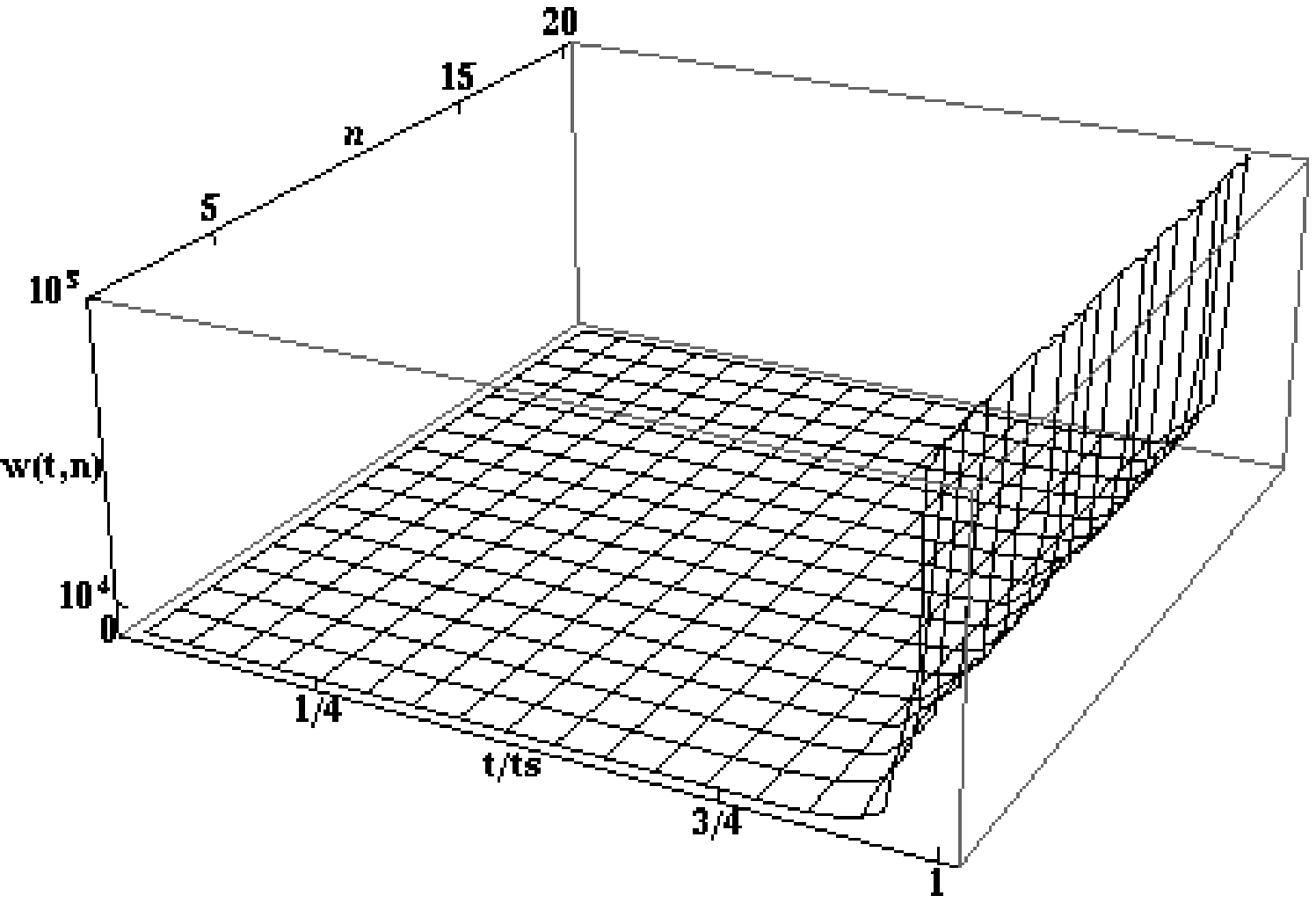}}
\label{w2}
\end{figure}

\section{Exotic-singularity-driven dark energy.}

\begin{figure}[h]
\label{SFS}
\caption{Distance modulus $\mu_L = m - M$ for the CC model with $H_0 = 72$kms$^{-1}$Mpc$^{-1}$, $\Omega_{m0} = 0.26$, $\Omega_{\Lambda 0} = 0.74$ (dashed curve) and
SFS model with $m=2/3, n=1.9999, \delta = -0.471, y_0 = 0.99936$ (solid
curve).}
\includegraphics[width=10cm,angle=270]{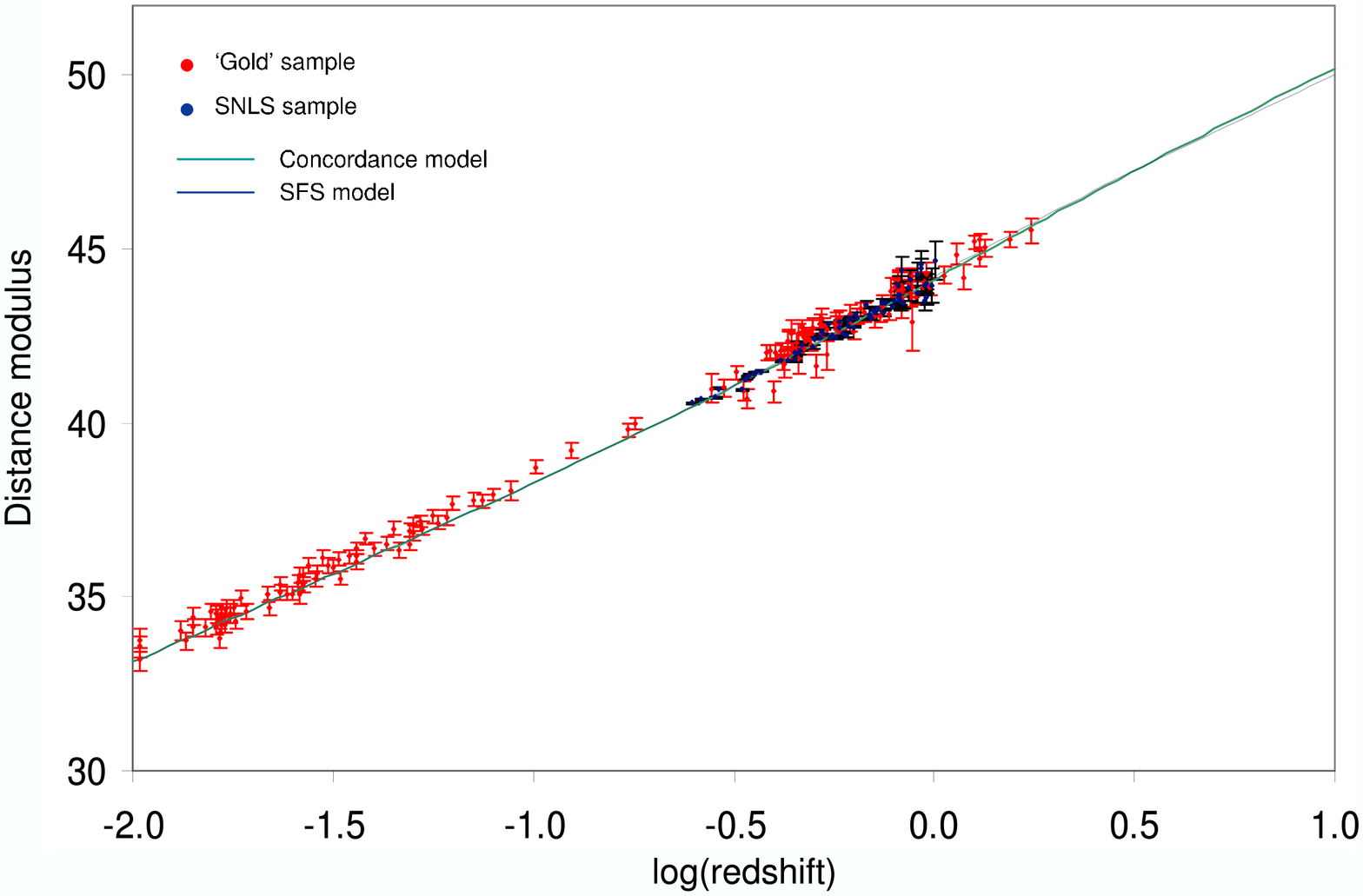}
\end{figure}

We have already mentioned that there is usually some fundamental physical theory (scalar field, higher-oder, string, brane, LQC) which can be related to the models with exotic singularities.
For example, a very peculiar $w-$singularity appears in physical theories such as $f(R)$ gravity, scalar field gravity, and in brane gravity \cite{starob80,saridakis,yuri}. On the other hand, SFS singularities  plague loop quantum cosmology \cite{wands}. In other words, the evidence for an exotic singularity may be attached to some form of matter which gives current acceleration of the universe and makes a candidate for the dark energy.
We now check which of these exotic singularity models can really serve that by {\it checking them against data} which favors accelerated universe. The best studied models are of course phantom models which still are within the range of observational limit. However, we will show that some other models (in particular SFS models) can play {\it a good candidate} for the dark energy, too. In order to do so we have compared SFS models with observational data and we have found that a SFS dark energy can mimic very well a $\Lambda$-term dark energy in the concordance cosmology (CC) \cite{SFS07}. We have shown that after taking the distance modulus $\mu_L = m - M$ as for the concordance (CC) model
($H_0 = 72$kms$^{-1}$Mpc$^{-1}$, $\Omega_{m0} = 0.26$, $\Omega_{\Lambda 0} = 0.74$)
and for an SFS model ($m=2/3, n=1.9999,$ $\delta = -0.471, y_0 = 0.99936$) both
models fit very well observational data (see Fig. \ref{SFS}).

Besides, we have come to a surprising remark. Namely, if the age of the SFS model is equal to the age of the CC model, i.e. $t_0=13.6$ Gyr, one finds that a SFS is possible in only 8.7 million years!

In this context, it is no wonder that the SFS singularities were termed ``sudden'', since they can really appear in a near future compared to a cosmological timescale. We have checked that a GSFS (a generalized SFS which leads to no energy conditions violation) are always more distant in future. This means the strongest of SFS type singularities is more likely to become reality.
A practical tool to recognize SFSs well in advance is to measure possible large values of statefinders \cite{snap,PLB05,AP06,ann06,dunaj} which are the higher-order characteristics of the universe expansion which go beyond the Hubble parameter and the deceleration parameter
$H = \dot{a}/a~$, $q  =  - \ddot{a}a/\dot{a}^2$~. They can generally be expressed as ($i \geq 2$) \cite{PLB05}
\bea
\label{dergen}
x^{(i)} &=& (-1)^{i+1}\frac{1}{H^{i}} \frac{a^{(i)}}{a} = (-1)^{i+1}
\frac{a^{(i)} a^{i-1}}{\dot{a}^{i}}~,
\eea
and the lowest order of them are known as:
jerk, snap (kerk), crack (lerk), pop (merk) etc.
\bea
\label{jerk}
j &=& \frac{1}{H^3} \frac{\dot{\ddot{a}}}{a} =
\frac{\dot{\ddot{a}}a^2}{\dot{a}^3}~, \hspace{0.5cm}
k = -\frac{1}{H^4} \frac{\ddot{\ddot{a}}}{a} = -\frac{\ddot{\ddot{a}}a^3}{\dot{a}^4}~,
\hspace{0.5cm} l = \frac{1}{H^5} \frac{a^{(5)}}{a} =
\frac{a^{(5)} a^4}{\dot{a}^5}~.
\eea
A blow-up of statefinders (possibly read-off a redshift-magnitude relation) may be linked to an emergence of future singularities as follows:
\be
\vert H_0 \vert \to \infty \Rightarrow {\rm BC, BR}, \hspace{0.2cm}
\mid q_0\mid \to \infty \Rightarrow {\rm SFS}, \hspace{0.2cm}
\mid j_0\mid \to \infty \Rightarrow {\rm GSFS}, \hspace{0.2cm}
{\rm etc.} \nonumber
\ee

\begin{figure}[h]
\label{FSF}
\caption{Distance modulus $\mu_L = m - M$ for the FSF model CC model with $\ddot{a} > 0$ 
for $\delta > 0$; $n = 0.87558$; $\delta$ = 0.53938; $t_0$ = 15.29571 Gyrs.}
\includegraphics[width=10cm,angle=0]{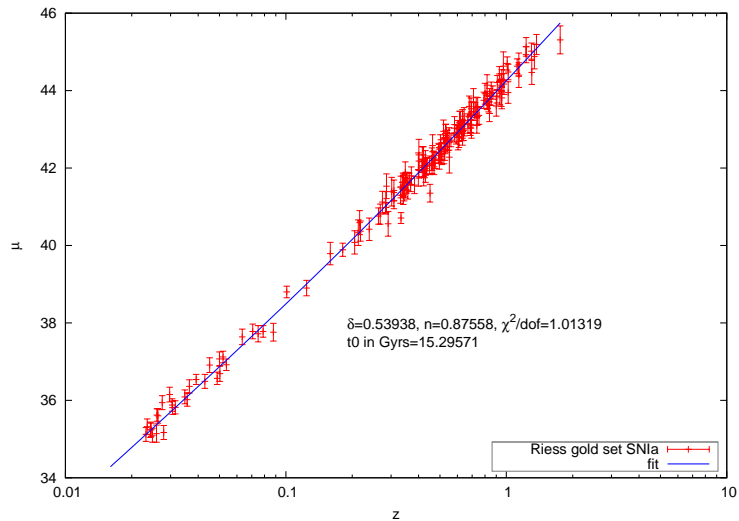}
\end{figure}

It is also interesting that an SFS with $a=a_b$ = const., $\dot{a} = 0$ ($\varrho \to 0$), and $\ddot{a} \to - \infty$ ($p \to \infty$) were also termed Big-Brake \cite{gorini}. They fulfill an anti-Chaplygin gas equation of state of the form
\be
p = \frac{A}{\varrho}\hspace{1.cm} {\rm A = const.}~~,
\ee
and this is the only case in which an explicit equation of state can be imposed
(since the relation allows $p \to \infty$ for $\varrho =$ const. $= 0$). They were studied
in the context of the tachyon cosmology in \cite{gergely}. However, due to the imposition of
different values of parameters which are given by tachyon constraints (plus anti-Chaplygin
gas constraints) the closest singularity in their model appears 1 Gyr in future, while the
furthest up to 44 Gyrs in future. Despite the distance, they of course can serve as a source
of the dark energy. We have also found that even the type III (Finite Scale Factor) singularity
can be closer than this, i.e. $t_s - t_0 \approx 0.3$ Gyrs (about 30 times larger than the
time to an SFS) with the choice of parameters to be:
$\ddot{a} > 0$ for $\delta > 0$; $n = 0.87558$; $\delta$ = 0.53938; $t_0$ = 15.29571 Gyrs (see
Fig. \ref{FSF}).

\section{Conclusions}

Our summarizing conclusions are as follows:

\begin{itemize}

\item Exotic singularities are related to new physical sources of
gravity which can {\it serve as the dark energy}.

\item First example source - phantom - produces an exotic singularity -- a Big-Rip (type I) in
which ($a \to \infty$ and $\varrho \to \infty$) and it is different from a Big-Bang/Big-Crunch.

\item Investigations of phantom {\it inspired} other searches for
  non-standard singularities (a Sudden Future Singularity (type II), a Generalized
  Sudden Future Singularity, a type III (a Finite Scale Factor), a type IV (a Big-Separation), a $w-$singularity) which, in fact, are not necessarily the ``true'' singularities (according to relativistic definitions), as sources of dark energy.

\item A Big-Rip and other exotic singularities are, in fact, {\it motivated by fundamental theories} of particle physics (scalar-tensor, superstring, brane, loop quantum cosmology etc.).

\item A Big-Rip which serves as dark energy may happen in 20 Gyr in future , while weak singularities (of tidal forces and their derivatives) may serve as dark energy if they are even quite close in the near future. For example an SFS may even appear in 8.7 Myr with no contradiction with data. A GSFS always appears later in future. Type III (FSF) is possible in about 0.3 Gyr. Finally, a Big-Brake (which is also an SFS) in tachyon cosmology context is at least 1 Gyr away from now.

\item Exotic singularities may manifest themselves in the higher-order characteristics of
expansion such as statefinders (jerk, kerk/snap, lerk/crack, merk/pop).
\end{itemize}





\begin{theacknowledgments}
We acknowledge the support of the Polish Ministry of Science
and Higher Education grant No N N202 1912 34 (years 2008-10).
\end{theacknowledgments}






\end{document}